\documentclass[A4paper, twocolumn]{article}
\usepackage{graphicx}
\usepackage{subfigure}
\usepackage{amsfonts,amssymb}
\usepackage{amsmath}
\usepackage{bbold}
\usepackage{dsfont}
\usepackage{color}
\usepackage{multicol}
\usepackage{abstract}
\usepackage{hyperref}

\begin{document}

\title{Identifying genuine entanglement of lossy noisy very large scale continuous variable Greenberger-Horne-Zeilinger state}

\author{Xiao-yu Chen \\ 
{\small {School of Information and Electrical Engineering, Hangzhou City University, Hangzhou {\rm 310015}, China }}}

\date{}

\twocolumn[
\maketitle

\begin{onecolabstract}


 Genuine entanglement identification of large scale systems is crucial for quantum computation, quantum communication and quantum learning advantage.  In contrast to experiments, where noisy intermediate-scale programmable photonic quantum processors have been developed, theoretically very limited results have been achieved for detecting genuine entanglement of continuous variable multipartite systems.  We propose a quite general and efficient entanglement detection framework for all kinds of multipartite entanglement of continuous variable systems based on uncertainty relations and the sign matrix technique.  Matrix criteria are demonstrated and can be  applied to various entanglement depth and k-separability problems of multimode systems. We illustrate the genuine entanglement conditions of continuous variable Greenberger-Horne-Zeilinger states of more than a hundred million modes in a photon loss and noise environment.  
\end{onecolabstract}
]

{\it Introduction}: Multimode continuous variable system has many applications in quantum computation\cite{Asavanant2019} and quantum communication\cite{Giovannetti}, it can also be used to achieve great quantum learning advantage\cite{Preskill}. Entanglement is the crucial resource in these applications. However, the certification of large scale genuine entanglement (a million mode cluster state in a time-multiplex manner \cite{Asavanant2019}) is not easy.

 Peres found the first entanglement criterion \cite{Peres} based on Werner's expression of separable quantum state \cite{Werner}. It is called positive partial transpose (PPT) criterion, namely, a state is necessarily separable if its partial transposed density matrix is positive, otherwise it is entangled.  The application of PPT criterion to continuous variable (CV) leads to an entanglement criterion for two-mode Gaussian states\cite{Simon}.  Meanwhile, uncertainty relation was proposed to build an entanglement criterion for the same two-mode Gaussian state system \cite{Duan}. Both results are necessary and sufficient and explicitly expressed with convariance matrix (CM) of the state. Further application of the PPT separable criterion shows that it is also necessary and sufficient for $1\times N$ Gaussian states but ceases to be sufficient for $2\times 2$ Gaussian states \cite{WernerWolf}, the authors demonstrated a CM entanglement criterion  which is necessary for all CV states and also sufficient for Gaussian states based on covariance matrix. The CM criterion is difficult to be directly apllied to detect entanglement since there are unknown subsystem covariance matrices yet to be determined. For multipartite system, the most common criteria are PPT criterion and van Loock-Furusawa (vLF) criterion\cite{Loock}\cite{Asavanant2024}. The genuine entanglement of quadripartite CV Greenberger-Horne-Zeilinger (GHZ) state and cluster state\cite{WangF}, hexapartite system\cite{JinJT2020}, octapartite system\cite{WangJW}\cite{Treps} are analyzed with vLF and PPT criteria.  However, vLF criterion can only detect multipartite CV entanglement for specific partitions\cite{Peng}. The biseparable set in vLF criterion adopted from \cite{Dur}\cite{Giedke} is not the convex hull of separable states with respect to any bipartitions\cite{Acin}. The later \cite{Acin} is a more popular and natural definition of biseparable state set\cite{Vicente}\cite{Makuta}.  When genuine entanglement is declared to be detected by vLF criterion, it may be a biseparable state yet inseparable under bipartions\cite{Guhne2010}. To our best knowledge, CV genuine entanglement criterion has also not been established based on quantum Fisher information\cite{Pezze}\cite{He}. Thus, new CV criterion should be proposed for CV genuine entanglement detection. 

We will follow Refs. \cite{Loock}\cite{Asavanant2024} to build the uncertainty relation using two linear combinations of position and momentum operators, but remove all the linear combination coefficients in vLF criterion to derive a matrix criterion called the sign matrix criterion based on CM. The convex hull of different partitions of the system is included in the criterion. The CM $\gamma$ of a CV quantum state $\rho$ is defined with its elements: $\gamma_{st}={\rm Tr}\{\rho[(\hat{\xi}_s-m_s)(\hat{\xi}_t-m_t)+(\hat{\xi}_t-m_t)(\hat{\xi}_s-m_s)]\}$, where $\hat{\xi}=(\hat{x}_1,\hat{x}_2,...,\hat{x}_n,\hat{p}_1,\hat{p}_2,...,\hat{p}_n)$ is the vector of operators and $\hat{x}_s,\hat{p}_s$ are the position and momentum operators of $s$-th mode, respectively. $m_s={\rm Tr}(\rho \hat{\xi}_{s})$ is the mean of $\hat{\xi}_s$. The Robertson-Schr\"{o}dinger uncertainty relation reads  $\gamma \geq i\Omega$, with $\Omega=\left(\tiny \begin{array}{cc}
   0 & -I \\
   I & 0 \\
 \end{array}
\underline{}\right)
 $, where I is the $n\times n$ identity matrix.
 
Consider an $n$-mode system with Hilbert space $H =H_1\otimes \cdot\cdot\cdot \otimes H_n$, the index set $\mathcal{J} =\{1, . . . , n\}$ are split into $k$ disjoint and nonempty subsets, denoted as $\mathcal{I}=\{\mathcal{I}_1,...,\mathcal{I}_k\}$. A quantum state $\rho_{\mathcal{I}}$ is called separable with respect to the special split $\mathcal{I}$ if it can be expressed as a mixture of  product states:
\begin{equation}\label{wee3}
\rho_{\mathcal{I}} =\sum_{i}p_i\rho_{\mathcal{I}_1}^{(i)}\otimes\rho_{\mathcal{I}_2}^{(i)}\otimes\cdot\cdot\cdot\otimes\rho_{\mathcal{I}_k}^{(i)}.
\end{equation}
Where $\{p_i\}$
is a probability distribution.  A quantum state $\rho$ is termed $\mathcal{D}$-separable if it can be expressed as:
\begin{equation}\label{wee5}
  \rho=\sum_{\mathcal{I}\in \mathcal{D}}q_{\mathcal{I}}\rho_{\mathcal{I}}.
\end{equation}
Where $\mathcal{D}$ is the set of $\mathcal{I}$ with some properties and $\{q_{\mathcal{I}}\}$
is a probability distribution.  If the number of subsets in $\mathcal{I}\in\mathcal{D} $ is fixed to be $K$, then $\rho$ is called $K$-separable. A biseparable state is with $K=2$.  A genuine entangled state is not biseparable. If the number of elements in subset $\mathcal{I}_i$ is denoted as $|\mathcal{I}_i|$, and let $J=\max_{i}|\mathcal{I}_i|$ be fixed for $\mathcal{I}\in\mathcal{D} $, then the state $\rho$ is called $J$-producible\cite{GuhneNJP}.

 {\it Sign matrix criterion framework}. Consider linear combinations $\hat{u}= \sum_{s}h_{s}\hat{\xi}_s$ and $\hat{v}=\sum_{s}g_{s}\hat{\xi}_s$ of position and momentum operators. The summation of the variances of $\hat{u}$ and $\hat{v}$ should be bounded from below according to the uncertainty relations. It is
\begin{equation}\label{wee1}
\langle(\Delta\hat{u})^2\rangle_{\rho}+\langle(\Delta\hat{v})^2\rangle_{\rho}\geq |\langle[\hat{u},\hat{v}]\rangle_{\rho}|=|h\Omega g^T|.
\end{equation}  
The inequality can be rewritten as $(h+ig)(\gamma\pm i\Omega)(h+ig)^{\dagger}\geq 0$, which leads to Robertson-Schr\"{o}dinger uncertainty relation. 
When the state $\rho$ is separable in the sence of (\ref{wee5}), we have 
\begin{equation}\label{we1}
\langle(\Delta\hat{u})^2\rangle_{\rho}+\langle(\Delta\hat{v})^2\rangle_{\rho}\geq\sum_{\mathcal{I}}q_{\mathcal{I}}\sum_{j}|h_{\mathcal{I}_j}\Omega_{\mathcal{I}_j}g_{\mathcal{I}_j}^T|,
\end{equation} 
where $h_{\mathcal{I}_j}$ is the subvector of $h$ with respect to index subset $\mathcal{I}_j$, and similar for $g$ and $\Omega$.

Proof of Eq. (\ref{we1}): 
Denote $\hat{u}_{\mathcal{I}_j}=\sum_{s\in\mathcal{I}_j}h_{s}\hat{\xi}_s$, then $\hat{u}=\sum_{j}\hat{u}_{\mathcal{I}_j}$.  We have 
\begin{eqnarray}\label{wee1b}
&&\langle(\Delta\hat{u})^2\rangle_{\rho}=\sum_{\mathcal{I}}q_{\mathcal{I}}\langle(\sum_{j}\hat{u}_{\mathcal{I}_j})^2\rangle_{\rho_{\mathcal{I}}}-\langle\hat{u}\rangle_{\rho}^2\nonumber \\
&&=\sum_{\mathcal{I}}q_{\mathcal{I}}\sum_{j}\langle(\Delta\hat{u}_{\mathcal{I}_j})^2\rangle_{\rho_{\mathcal{I}}}+\sum_{\mathcal{I}}q_{\mathcal{I}}(\sum_{j}\langle\hat{u}_{\mathcal{I}_j}\rangle_{\rho_{\mathcal{I}}})^2-\langle\hat{u}\rangle_{\rho}^2\nonumber\\
&&=\sum_{\mathcal{I}}q_{\mathcal{I}}\sum_{j}\langle(\Delta\hat{u}_{\mathcal{I}_j})^2\rangle_{\rho_{\mathcal{I}}}+\sum_{\mathcal{I}}q_{\mathcal{I}}(\langle\hat{u}\rangle_{\rho_{\mathcal{I}}})^2\nonumber\\
&&-(\sum_{\mathcal{I}}q_{\mathcal{I}}\langle\hat{u}\rangle_{\rho_{\mathcal{I}}})^2\geq\sum_{\mathcal{I}}q_{\mathcal{I}}\sum_{j}\langle(\Delta\hat{u}_{\mathcal{I}_j})^2\rangle_{\rho_{\mathcal{I}}}.
\end{eqnarray}  
Where we have applied the Cauchy-Schwarz inequality $(\sum_{\mathcal{I}}q_{\mathcal{I}})(\sum_{\mathcal{I}}q_{\mathcal{I}}(\langle\hat{u}\rangle_{\rho_{\mathcal{I}}})^2)\geq (\sum_{\mathcal{I}}q_{\mathcal{I}}\langle\hat{u}\rangle_{\rho_{\mathcal{I}}})^2$. 
Furthermore we have 
\begin{eqnarray}\label{wee1c}
\langle(\Delta\hat{u}_{\mathcal{I}_j})^2\rangle_{\rho_{\mathcal{I}}}=\sum_{i}p_{i}\langle\hat{u}_{\mathcal{I}_j}^2\rangle_{\rho_{\mathcal{I}_j}^{(i)}}-\langle\hat{u}_{\mathcal{I}_j}\rangle_{\rho_{\mathcal{I}}}^2\nonumber\\
=\sum_{i}p_{i}\langle(\Delta\hat{u}_{\mathcal{I}_j})^2\rangle_{\rho_{\mathcal{I}_j}^{(i)}}+\sum_{i}p_{i}(\langle\hat{u}_{\mathcal{I}_j}\rangle_{\rho_{\mathcal{I}_j}^{(i)}})^2\nonumber\\-(\sum_{i}p_{i}\langle\hat{u}_{\mathcal{I}_j}\rangle_{\rho_{\mathcal{I}_j}^{(i)}})^2\geq\sum_{i}p_{i}\langle(\Delta\hat{u}_{\mathcal{I}_j})^2\rangle_{\rho_{\mathcal{I}_j}^{(i)}}. 
\end{eqnarray}  
Where we have used the Cauchy-Schwarz inequality $(\sum_{i}p_{i})(\sum_{i}p_{i}(\langle\hat{u}_{\mathcal{I}_j}\rangle_{\rho_{\mathcal{I}_j}^{(i)}})^2)\geq(\sum_{i}p_{i}\langle\hat{u}_{\mathcal{I}_j}\rangle_{\rho_{\mathcal{I}_j}^{(i)}})^2$. Thus it leads to
\begin{equation}\label{wee1a}
\langle(\Delta\hat{u})^2\rangle_{\rho}\geq \sum_{\mathcal{I}}q_{\mathcal{I}}\sum_{j}\sum_{i}p_{i}\langle(\Delta\hat{u}_{\mathcal{I}_j})^2\rangle_{\rho_{\mathcal{I}_{j}}^{(i)}}.
\end{equation}  
Applying the uncertainty relation (\ref{wee1}) to the subset $\mathcal{I}_{j}$, we have $\langle(\Delta\hat{u}_{\mathcal{I}_j})^2\rangle_{\rho_{\mathcal{I}_{j}}^{(i)}}+\langle(\Delta\hat{v}_{\mathcal{I}_j})^2\rangle_{\rho_{\mathcal{I}_{j}}^{(i)}}\geq|h_{\mathcal{I}_j}\Omega_{\mathcal{I}_j}g_{\mathcal{I}_j}^T|.$
We then get the uncertainty relation (\ref{we1}) for a separable state defined in (\ref{wee5}).  

 Clearly, for a separable state, the bound in (\ref{we1}) is tighter than that given in (\ref{wee1}). Our mission is to remove the absolute signs in (\ref{we1}), such that the vectors $h$ and $g$ can be eliminated and the inequality becomes a matrix inequality. To remove an absolute sign in the bound, there are two possible results (namely $\pm$), so the number of possible results increase rapidly. We may write  $\sum_{j}|h_{\mathcal{I}_j}\Omega_{\mathcal{I}_j}g_{\mathcal{I}_j}^T|$ as  $h\tilde{\Omega}g^T$, where $\tilde{\Omega} $ is derived from $\Omega$ by changing the sign of some nonzro elements. At last, a matrix creterion can be obtained for the different kinds of separable states:
\begin{equation}\label{wee11}
\gamma +\sigma_2\otimes Q \geq 0.
\end{equation}
Where $\sigma_2$ is the second Pauli matrix, and $Q$ is a diagonal matrix, its diagonal elements are determined by the product of probability distribution $\{q_{\mathcal{I}} \}$ and a matrix $T$ called the sign matrix, with its elements are $\pm1$. 

We will use three-mode CV state to illustrate the process of removing $h,g$ vectors of vLF criterion\cite{Loock} in deriving sign matrix criterion. The biseparable state takes the following form
\begin{equation}\label{wee100}
    \rho_{\rm bisep}=q_{1}\rho_{1|23}+q_{2}\rho_{2|13}+q_{3}\rho_{12|3}.
\end{equation}
Where $\rho_{1|23}=\rho_{1}\otimes\rho_{23}$ and so on. The uncertainty relations for the biseparable state is
\begin{equation}\label{wee101}
(\langle(\Delta\hat{u})^2\rangle+\langle(\Delta\hat{v})^2\rangle)_{\rho_{\rm bisep}}\geq f(\bold{q},\bold{b}).
\end{equation}
 where $f(\bold{q},\bold{b})=q_1( |b_{1}|+|b_{2}+b_{3}|)+q_2(|b_{2}|+|b_{1}+b_{3}|)+q_3(|b_{3}|+|b_{1}+b_{2}|)$, with $\bold{q}=(q_1,q_2,q_3),\bold{b}=(b_1,b_2,b_3)$ and $b_{i}=h_{i}g_{i+3}-h_{i+3}g_{i}$. Without loss of generality, let $|b_{1}|\ge|b_{2}|\ge|b_{3}|$ and $b_{1}>0$, we have several cases:(1) $b_{2}<0, b_3<0$, (2) $b_{2}<0,b_{3}>0$, (3)$b_{2}>0,b_{3}<0$. Case (1) leads to $f(\bold{q},\bold{b})=q_1( b_{1}-b_{2}-b_{3})+q_2(b_{1}-b_{2}+b_{3})+q_3(b_{1}+b_{2}-b_{3})=\bold{q}T_{1}\bold{b}^T$, similarly, Cases (2) and (3) lead to $f(\bold{q},\bold{b})=\bold{q}T_{2}\bold{b}^T$ and $\bold{q}T_{3}\bold{b}^T$, with  $T_1,T_2,T_3$ being 
\begin{equation}\label{wee14}
 T_{1}=\begin{bmatrix}
   1 & -1& -1\\
  1 &-1&1  \\
  1&1&-1\\
 \end{bmatrix},
 T_{2}=\begin{bmatrix}
   1 & -1& -1\\
  1&-1&1  \\
  1&1&1\\
 \end{bmatrix},
T_{3}=\begin{bmatrix}
   1 & 1& 1\\
  1&1&1  \\
  1&1&-1\\
 \end{bmatrix},
\end{equation}
respectively.  We may call $T_{i}$ sign matrix since its elements are either $1$ or $-1$. More sign matrices can be obtained by mode permutations. There are totally 12 sign matrices for the biseparability of three mode system. Since $Q$ is diagonal, we denote $V_{i}=Q_{ii}$. The vector $\bold{V}=\bold{q} T$, where $T$ is one of the 12 different sign matrices. 

For an arbitary three mode state with its CM, the criterion (\ref{wee11}) can be applied to give a more explicit biseparable condition by optimizing the probability distribution. Let the probability distribution be $\{q_1,q_2,q_3\}$ as shown in (\ref{wee100}) , for each randomly chosen probability sample, we calculate the smallest eigenvalue of  matrix at the left hand side of (\ref{wee11}) for every $T_{i}$, then find the minimal one over all $T$ matrices, and record a value. We then calculate  the values for the randomly chosen probability distribution samples many times. If the maximum of these values is still negative, inequality (\ref{wee11}) is violated for the whole probability distribution, the state is genuinely entangled, otherwise the state is biseparable. So the process is to minimize the smallest eigenvalue of  the matrix in (\ref{wee11}) over the set of sign matrices, then maximize the results with respect to probability distribution. If the final result is still negative, then the state is genuinely entangled.

\begin{figure}
\centering
\subfigure[\label{Fig.1a}]{
\includegraphics[width=1.65in]{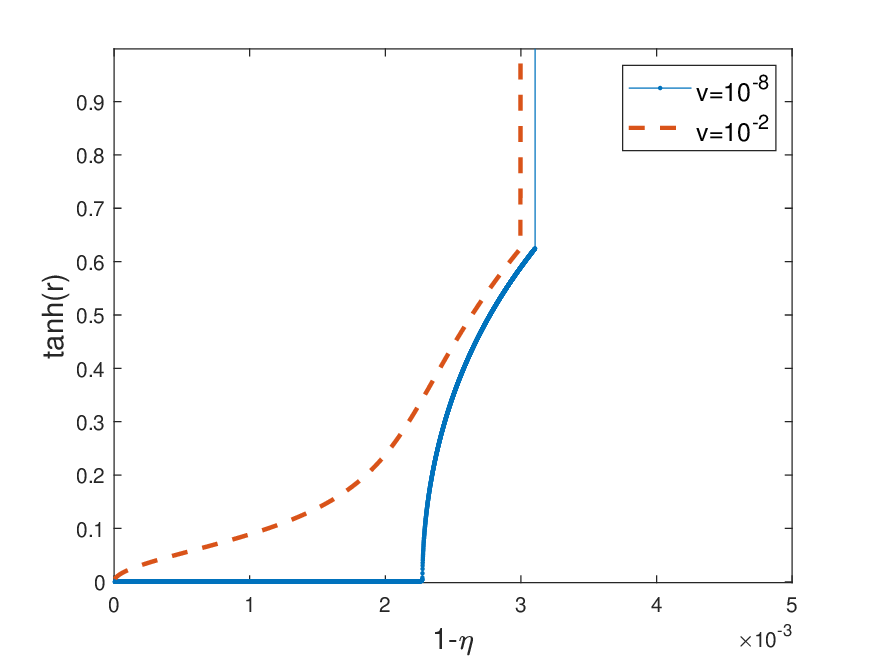}}
\subfigure[\label{Fig.1b}]{
\includegraphics[width=1.65in]{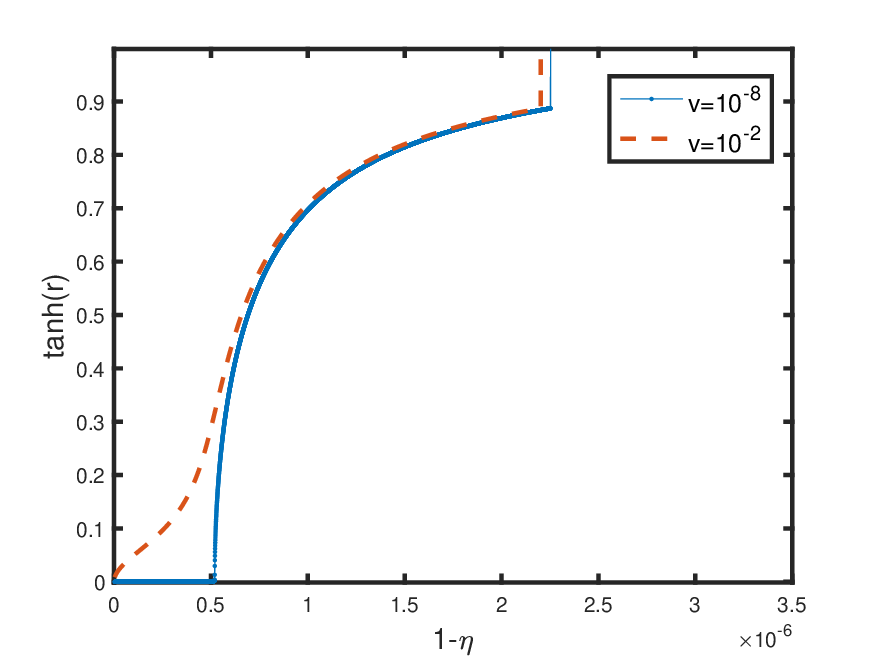}}
\subfigure[\label{Fig.1c}]{
\includegraphics[width=1.65in]{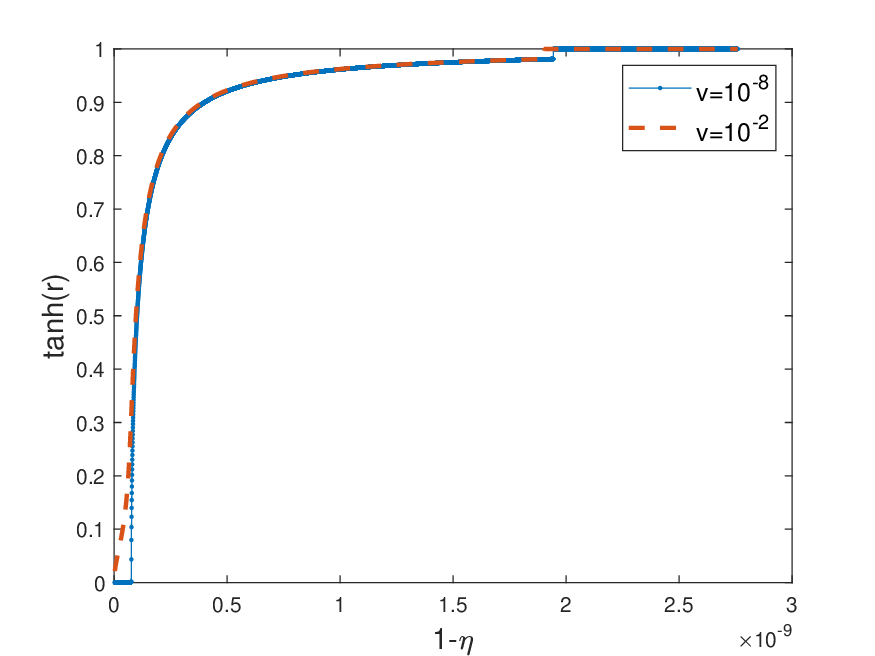}}
\subfigure[\label{Fig.1d}]{
\includegraphics[width=1.65in]{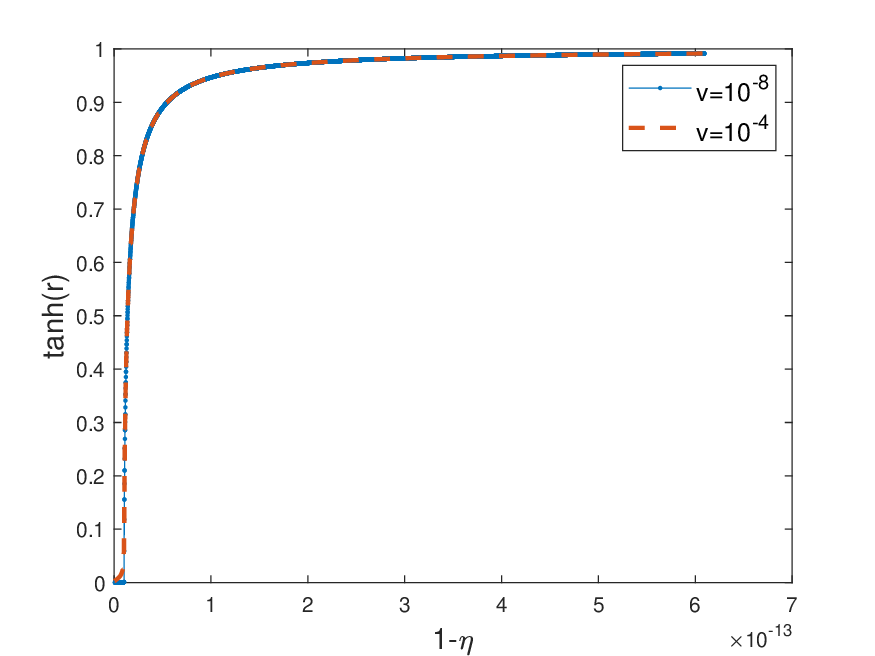}}
\caption{ Genuine entanglement detection of lossy noisy CV-GHZ states, where $v=N/(N+1)$. The longitudinal straight lines in the subfigures are due to the additional condition. (a) $n=10^2$ modes, with $m=6, n_0=1$.  (b) $n=10^4$ modes, with $m=13, n_0=1$.  (c) $n=10^6$ modes, with $m=20, n_0=500000$. (d) $n=10^8$ modes, with $m=26, n_0=1$.}
\label{Fig1}
\end{figure}
 
If we do not know the whole family of sign matrices, we only have one known sign matrix, we can still get a necessary condition of biseparability, which can detect some of genuine entanglement.

{\it Genuine entanglement identification of massive mode noisy CV-GHZ state}.
The application of the sign matrix criterion to continuous-variable GHZ state in loss and noise environment will give a necessary condition of biseparability.  A lossy and noisy continuous variable GHZ state is characterized by its CM $\gamma=\gamma_{x}\oplus\gamma_{p}$, with all the diagonal elements of $\gamma_{x}$ are $a$, off-diagonal elements are $c$, all the diagonal elements of $\gamma_{p}$  are $b$ and off-diagonal elements are $-c$. Where $a=\eta a_0+N_1$,$b=\eta b_{0}+N_1$, $c=\eta c_{0}$ and $N_1=(1-\eta)(2N+1)$\cite{Giovannetti2012}. The loss is described by transmissivity $\eta$, the average photon number of noise is denoted as $N$. The initial pure CV-GHZ state is described by $a_0=\frac{1}{n}(e^{2r}+(n-1)e^{-2r})$,$b_0=\frac{1}{n}((n-1)e^{2r}+e^{-2r})$, $c_0=\frac{1}{n}(e^{2r}-e^{-2r})$, where the number of modes is $n$, and the squeezing parameter is $r$. For a CM of the form $\gamma=\gamma_{x}\oplus\gamma_{p}$, the sign matrix criterion (\ref{wee11}) can be equivalently written as  
\begin{equation}\label{wee11a}
\left( \begin{array}{cc}
   \gamma_x& Q\\
  Q& \gamma_p \\
 \end{array}
\right)\geq 0.
\end{equation}
The diagonal matrix Q takes the form of $diag(Q)=(1,1,...,1,\kappa,\kappa,...,\kappa)$. This assumption will be proven later. The number of $1$ in $diag(Q)$ is $m$, the number of $\kappa$ is $n-m$, where $\kappa$ is a function of $n,m$ and $n_0$ when the system is split into $n_0$ modes and $n-n_0$ modes in the bipartition. Due to symmetry, the criterion (\ref{wee11a}) reduces to the following inequality
\begin{equation}\label{wee11b}
\left( \begin{array}{cccc}
  a'& c'&1&0\\
   c'& b'&0&\kappa\\
  1&0&b'&-c'\\
  0&\kappa&-c'&b'\\
 \end{array}
\right)\geq 0.
\end{equation}
Where $a'=a+(m-1)c$, $b'=b-(m-1)c$, $c'=\sqrt{(n-m)m}c$. The positivity of the determinant of the matrix in (\ref{wee11b}) leads to a quadratic equation of $\cosh(2r)$ and the necessary condition for the biseparability can be explicitly expressed. We show the results of sufficient conditions for $10^2, 10^4,10^6,10^8$ modes of the lossy noisy CV-GHZ states to be genuinely entangled in Figure \ref{Fig1}. There is an additional condition in the application of the criterion to massive mode CV-GHZ states, it is expressed with the eigenvector of matrix in (\ref{wee11b}), and will be explained later.

\begin{figure}
\centering
\subfigure[\label{Fig.2a}]{
\includegraphics[width=1.65in]{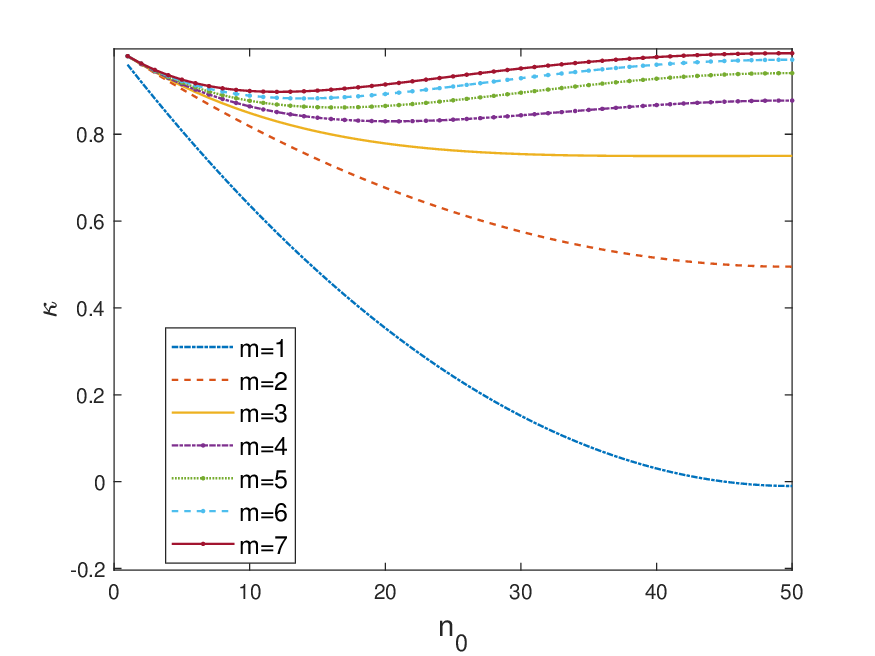}}
\subfigure[\label{Fig.2b}]{
\includegraphics[width=1.65in]{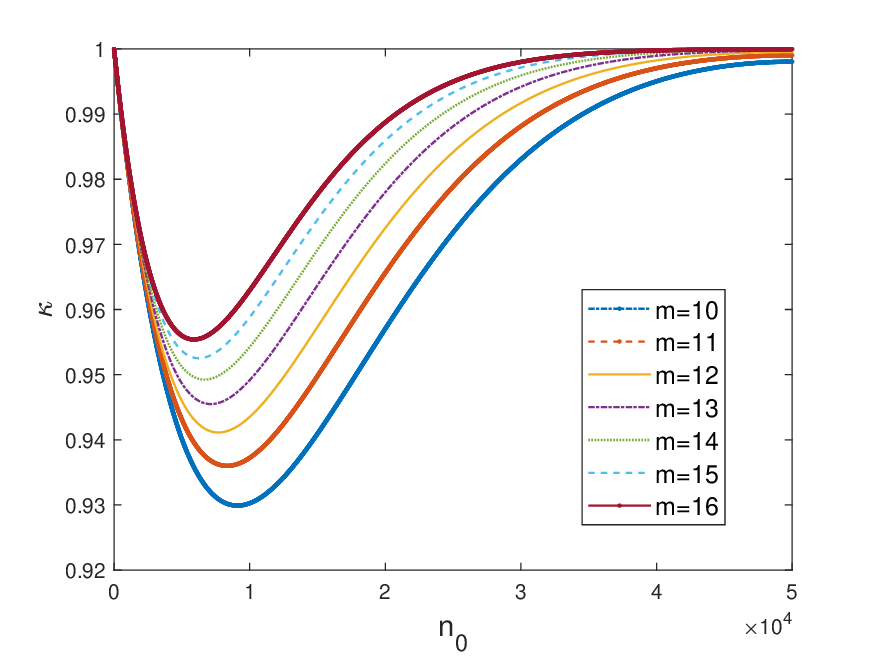}}
\caption{ (a) (b) $\kappa$ in Eqs. (\ref{wee11a}) and (\ref{wee11b}) as a function of $n_{0}$ for $n=10^2$ and $n=10^5$ mode systems. }
\label{Fig2}
\end{figure}

For an $n$-mode lossy and noisy CV-GHZ state with CM described as in former section, we consider the bisepartition with $n_0$ modes in one party and $n-n_0$ modes in another party. Let $\bold{b}=(1,1,...1,-\tau,-\tau,...,-\tau)$, the number of $1$ in $\bold{b}$ is $m$, the number of $-\tau$ in $\bold{b}$ is $n-m$. Given the number $n_0$, we have $C_n^{n_0}$ (combinatorial number) splits that devide the system into two parts, each split may be assigned with a probability $q_{i}$, with $i=1,2,...,C_n^{n_0}$. When $n_0<m$, we have two different cases of choosing $n_0$ modes from all $n$ modes: (i). Choose at least one mode from the first $m$ modes, we have a  $(C_n^{n_0}-C_{n-m}^{n_0})\times n$ marix with all elements being $1$ as the first submatrix of sign matrix $T$ as far as
\begin{equation}\label{we12}
(n-m)\tau<1.
\end{equation}
(ii). Choose all $n_{0}$ modes from the last $n-m$ modes, we get $(1,1,...1,-1,-1,...,-1,1,1,...,1)$ as a line of sign matrix if the $m+1,m+2,...,m+n_0$ modes are chosen. The sign matrix has a submatrix of size $C_{n-m}^{n_0}\times n$ with $n_0$ number of $-1$ at the last $n-m$ columns in each line. The position of $-1$ relies on which mode is chosen. Thus the sign matrix can be determined. Similarly, when $n_0\geq m$, we have three different cases of choosing $n_0$ modes from all $n$ modes, the  
sign matrix can also be easily determined. 

With the known sign matrix $T$, we proceed to the diagonal matrix $Q$, we have $\bold{V}=diag(Q)=\bold{q}T$.  We need to find the optimal probability distribution $\bold{q}$ such that the minimal eigenvalue of  $\gamma_x\oplus\gamma_p+\sigma_1\otimes Q$ in the left hand side of (\ref{wee11a}) is maximized with respect to $\bold{q}$, where $\sigma_1$ is the first Pauli matrix. For any probability distribution $\bold{q}$, we have $\bold{V}=(1,1,...,1,\kappa_{1},\kappa_{2},...,\kappa_{n-m})$. The first $m$ components of $\bold{V}$ are 1 due to the fact that the left block of T is a $C_n^{n_0}\times m$ submatrix with all elements being 1. $\kappa_{i}=\sum_jq_jT_{m+i,j}$. The average $\kappa=\frac{1}{n-m}\sum_i\kappa_i$ is determined to be 

\begin{equation} \label{wee11d}
 \kappa= \begin{cases} 1-2\frac{C_{n-m-1}^{n_0-1}}{C_{n}^{n_0}}, \quad \quad \quad\text{if}\quad   n_0<m;\\
 1-2\frac{(C_{n-m-1}^{n_0-1}+C_{n-m-1}^{n_0-m})}{C_{n}^{n_0}}, 
  \quad \quad   \text{otherwise}.
\end{cases} 
 \end{equation}
via counting the number of $-1$ in sign matrix $T$, as shown in Figures \ref{Fig.2a} and \ref{Fig.2b}.

In order to discriminate different $Q$ matrices, we denote $Q_{\kappa}$ with its diagonal vector  $\bold{V}_{\kappa}=(1,1,...,1,\kappa,\kappa,...,\kappa)$ and $Q$ with its diagonal vector $\bold{V}=(1,1,...,1,\kappa_{1},\kappa_{2},...,\kappa_{n-m})$. Let the eigenvector coresponding to the minimal eigenvalue of  matrix $\gamma_x\oplus\gamma_p+\sigma_1\otimes Q(\kappa)$ be $|\psi\rangle=|(\alpha,\alpha,...,\alpha,-\beta,-\beta,...,-\beta,\delta,\delta,...,\delta,\omega,\omega,...,\omega)^T\rangle$, where $\alpha,\beta,\delta,\omega$ are real (actually they are positive), the numbers of $\alpha$ and $\delta$ in $\psi$ are $m$, the numbers of $-\beta$ and $\omega$ are $n-m$. We have $\langle\psi|\sigma_1\otimes Q|\psi\rangle=2(m\alpha\delta-\beta\omega\sum_{i=1}^{n-m}\kappa_{i})=\langle\psi|\sigma_1\otimes Q_{\kappa}|\psi\rangle$. Hence
\begin{equation} \label{wee11e}
\langle\psi|\gamma_x\oplus\gamma_p+\sigma_1\otimes Q|\psi\rangle=\langle\psi|\gamma_x\oplus\gamma_p+\sigma_1\otimes Q_{\kappa}|\psi\rangle.
\end{equation}
Denote the minimal eigenvalue of matrix A to be $\Lambda(A)$, we have 
\begin{equation} \label{wee11f}
\Lambda\left( \begin{array}{cc}
   \gamma_x& Q\\
  Q& \gamma_p \\
 \end{array}
\right)\leq \langle\psi|\left( \begin{array}{cc}
   \gamma_x& Q\\
  Q& \gamma_p \\
 \end{array}
\right)|\psi\rangle=\Lambda\left( \begin{array}{cc}
   \gamma_x& Q_{\kappa}\\
  Q_{\kappa}& \gamma_p \\
 \end{array}
\right).
\end{equation}
The inequality comes from the fact that the mean of a matrix on an arbitary vector is larger than its minimal eigenvalue. The equality comes from (\ref{wee11e}) and $|\psi\rangle$ is the eigenvector corresponding to the minimal eigenvalue of $\gamma_x\oplus\gamma_p+\sigma_1\otimes Q_{\kappa}$. Hence we show that $\kappa_{1}=\kappa_{2}=...=\kappa_{n-m}=\kappa$ achieves the largest minimal eigenvalue of $\gamma_x\oplus\gamma_p+\sigma_1\otimes Q$ among all valid $\boldsymbol{\kappa}$. With mode permutations, it is easy to show  that a uniform probability distribution $\bold{q}=(1,1,...,1)/C_{n}^{n_0}$ is the right solution. 

Next, we consider the problem of genuine entanglement region changing with repect to $\kappa$, we will show that the genuine entanglement range expands with the increase of $\kappa$. Denote the determinant of the left hand side of (\ref{wee11b}) as $y(\kappa)$, we have 
\begin{equation}
y(\kappa)=(a'b'-c'^2)^2-(1+\kappa^2)a'b'+2\kappa c'^2+\kappa^2.
\end{equation}
It is known that $a'b'>1$, so $y(\kappa)$ is a downwards parabola. Let the zero points of $y(\kappa)$ be $\kappa_{\mp}$, since $y(1)=(a'b'-c'^2-1)^2>0$, we have $\kappa_{+}>1>\kappa_{-}$. From (\ref{wee11d}), we know that $\kappa<1$. So we drop $\kappa_{+}$ zero point and keep $\kappa_{-}$ zero point. We have $\frac{dy}{d\kappa}|_{\kappa=\kappa_{-}}>0$.  This means that $y(\kappa_{-}+\Delta\kappa)>y(\kappa_{-})=0$, so the border line of genuine entanglement and necessarily biseparable moves towards genuine entanglement side, the region of genuinely entangled state set decreases when $\kappa$ increases. The largest region of biseparability occurs when $\kappa$ is maximized.  In Figure \ref{Fig.2b}, $\kappa$ as function of $n_0$ for $n=100$ mode and $n=100000$ systems with typical $m$ is shown. The maximal $\kappa$ appears when $n_0=1$ or $n_0=\lfloor \frac{n}{2} \rfloor$. Notice that a mix of states with different $n_{0}$ does not increase the maximal $\kappa$. The region of biseparability for maximal $\kappa$ covers the region of biseparability for all the other smaller $\kappa$. So we do not need to consider the possibility of  
the mix of different $n_0$ partitions.

For the genuine entanglement identification of an $n$-mode lossy and noise CV-GHZ state, we choose the number $m$ such that $\kappa(n_0=1)$ and $\kappa(n_0=\lfloor \frac{n}{2} \rfloor)$ are comparable. Then we choose the larger one of $\kappa(n_0=1)$ and $\kappa(n_0=\lfloor \frac{n}{2} \rfloor)$ for the genuine entanglement condition. The genuine entanglement region is minimized with respect to $n_0$. Then we use different values of $m$ to maximize the genuine entanglement region  with respect to $m$.

Finally, the additional condition mentioned above comes from (\ref{we12}). The eigenvector corresponding to the minimal eigenvalue of $\gamma_x\oplus\gamma_p+\sigma_1\otimes Q_{\kappa}$ is $|\psi\rangle$. From $|\psi\rangle$ we obtain $\bold{h}=(\alpha,\alpha,...,\alpha,-\beta,-\beta,...,-\beta,0,0,...,0)$, the number of $\alpha$ is $m$, the number of $\beta$ is $n-m$ and the number of $0$ is $n$. Similarly, $\bold{g}=(0,0,...,0,\delta,\delta,...,\delta,\omega,\omega,...,\omega)$. Hence $\bold{b}=(\alpha\delta,\alpha\delta,...,\alpha\delta,-\beta\omega,-\beta\omega,...,-\beta\omega)$, comparing with the unnormalized version of $\bold{b}=(1,1,...1,-\tau,-\tau,...,-\tau)$ introduced before (these two versions of $\bold{b}$ are up to a constant factor), we have $\tau=\frac{\beta\omega}{\alpha\delta}$. Denote the eigenvector corresponding to the minimal eigenvalue of matrix at the left hand of (\ref{wee11b}) as $(\alpha',-\beta',\delta',\omega')^T$, we have $\alpha'=\sqrt{m}\alpha, \beta'=\sqrt{n-m}\beta,\delta'=\sqrt{m}\delta,\omega'=\sqrt{n-m}\omega$. Hence the additional condition is
\begin{equation}\label{we12a}
m\frac{\beta'\omega'}{\alpha'\delta'}<1.
\end{equation}

For $n$ from $3$ to $129247013$, we list the parameters $m$ and $n_0$ in the Supplimental Material. The condition of biseparability is also shown in a directly calculable fashion. 

{\it Discussion}. The method developed in this paper is also valid for all the other kinds of separabilities. The only difference is that we should replace the sign matrix for the general k-separable problems.  We list the results for $n\leq 1029$ in the Supplemental Material for triseparble and quadriseparable cases of CV-GHZ state. We provide some results in the Supplemental Material for J-producibilities of lossy noisy multimode CV-GHZ states.
 
In summary, we have proposed a framework of multipartite entanglement detection for continuous variable systems. It corresponds to a witness which is a quadratic form of position and momentum operators with variable combination parameters. The validity of the witness comes from uncertainty relations. The optimization of all the combination parameters leads to strongly optimal (or matched) entanglement witness. Multipartite entanglement detecting problem is transformed to the negetivity of a matrix which is the CM plus a sparse matrix determined by probability distribution and a sign matrix. The sign matrix characterizes the partition manner of a state to be split into different parties.  All different kinds of multipartite entanglement share the same form of criterion, with different sign matrices. The criterion is scalable and we have presented genuine entanglement conditions for lossy and noisy very large scale CV-GHZ states up to a hundred million modes.  

This work is supported by the National Natural Science Foundation  of China (Grant No.61871347)

\bibliographystyle{unsrt}
\bibliography{SMECbibfile}

\newpage

\setcounter{equation}{0}
\renewcommand\theequation{S.\arabic{equation}} 

\section*{Supplemental material: }
\subsection*{A. Details of genuine entanglement conditions from 3 to 129247013 modes }
The necessary biseparable condition is 
\begin{equation}\label{S1}
(a'b'-c'^2)^2-(1+\kappa^2)a'b'+2\kappa c'^2+\kappa^2\geq 0.
\end{equation}
Since $a'b'-c'^2=\eta^2+N_1^2+2\eta N_1\cosh(2r)$ is a linear function of $\cosh(2r)$, the inequality (\ref{S1}) is a quadratic inequality of $\cosh(2r)$, it is
\begin{equation}\label{S2}
A\cosh^2(2r)+B\cosh(2r)+C\geq 0.
\end{equation}
Where
\begin{eqnarray}\label{S3}
&&A=4\eta^2(N_{1}^2-(1-\kappa)^2(n-m)m/n^2),\nonumber\\
&&B=4\eta N_1(\eta^2+N_1^2)-2\eta N_1(1+\kappa^2),\nonumber\\
&&C=(\eta^2+N_1^2-1)(\eta^2+N_1^2-\kappa^2)\nonumber\\
&&+4\eta^2(1-\kappa)^2(n-m)m/n^2.\nonumber
\end{eqnarray}
\begin{table}[h!]
\begin{center}
\caption{Parameters of biseparable condition.}
\label{TTTT1}
\begin{tabular}{l|c|c} 
\hline
\text{m} & \text{n} & \text{n}\\
 & $n_1=n-\lfloor \frac{n}{2} \rfloor$ & $n_1=n-1$ \\
 & $n_0=\lfloor \frac{n}{2} \rfloor$ & $n_0=1$ \\
\hline

1 &  & 3-9\\
3 & & 10-16\\
4 & 17-18&19-32 \\
5 & 33-37&38-63\\
6 &64-73&74-124\\
7 &125-142&143-244\\
8 & 245-275&276-482 \\
9 & 483-538&539-957\\
10&958-1058&1059-1906\\
11 &1907-2091&2092-3809\\
12 &3810-4149&4150-7627 \\
13 &7628-8256&8257-15291\\
14&15292-16460&16461-30674\\
15&30675-32857&32858-61544\\
16& 51545-65639&65640-123481 \\
17 & 123482-131190&131191-247715\\
18& 247716-262278& 262279-496848\\
19& 496849-524439&524440-996318\\
20 & 996319-1048745&1048746-1997477 \\
21 &1997478-2097340&2097341-4003864\\
22 &4003865-4194512&4194513-8024117\\
23 &8024118-8388838& 8388839-16078419\\
24 &16078420-16777468&16777469-32212527 \\
25 & 32212528-33554707&33554708-64528051\\
26 & 64528052-67109163&67109164-129247012\\
27 & 129247013& \\

\hline
\end{tabular}
\end{center}
\end{table}
Hence, the necessary condition of biseparability of CV-GHZ states can be analytically solved. The result is that the squeezing parameter $r$ is an explicit function of transmissivity parameter $\eta$ and average noise photon number $N$.  In Table.\ref{TTTT1},the $n$ mode state is split into two parts, one has $n_0$ modes and the other has $n_1=n-n_0$ modes. For a given $n_0$, we have $C_{n}^{n_0}$ different ways to choose $n_0$ modes from $n$ modes. All the $C_{n}^{n_0}$ choices have the same probability $1/C_{n}^{n_0}$ for the CV-GHZ states as shown in the main paper. For a given $m$, we find the optimal $n_0$ such that the genuine entanglement range is minimized. The reason of minimization of the genuine entanglement range over $n_0$ is that a genuine entanglement state should not be decomposed to biseparble state for any partition. Notice that the number $m$ specifies the witness. For different witnesses, we should compare them to find a best one which detects as many genuine entanglement states as possible. Thus, for given $m$ in Table.\ref{TTTT1}, we compare different $n_0$ to minimize the genuine entanglement range. For different $m$, we compare them to maximize the genuine entanglement range.

\subsection*{B. Parameters for parts of K-separabilities and J-producibilities for CV-GHZ states}

\begin{figure}
\centering
\subfigure[\label{Fig.3a}]{
\includegraphics[width=1.65in]{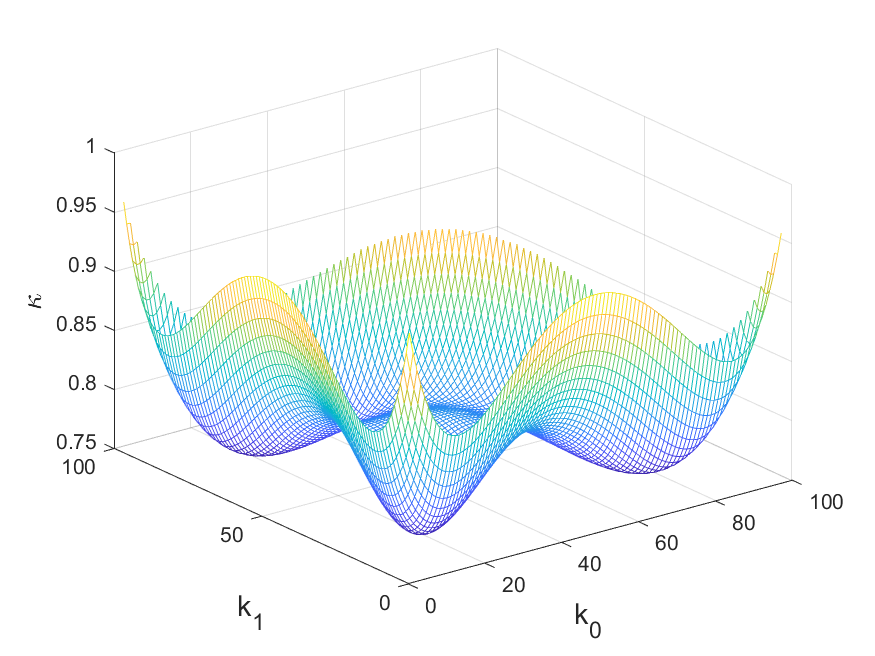}}
\subfigure[\label{Fig.3b}]{
\includegraphics[width=1.65in]{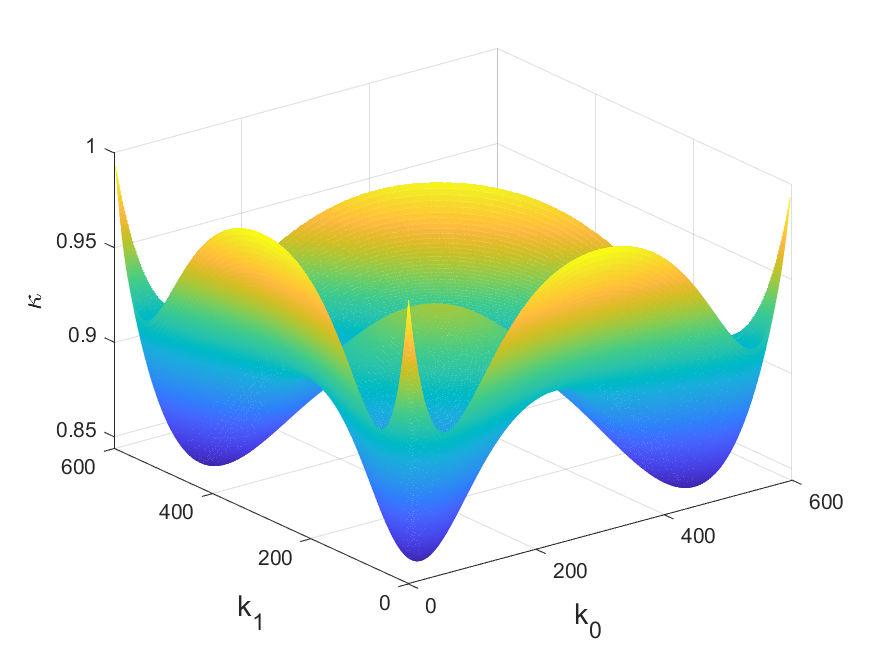}}
\caption{(a)(b)  $\kappa$ in the triseparable case as a function of $n_0$ and $n_1$ for $n=100$ mode system with $m=6$ and $n=600$ system with $m=9$, respectively. There are $n_0,n_1,n-n_0-n_1$ modes in the three parts of the tripartite partition.}
\label{Fig3}
\end{figure}

For the tripartite separability (triseparability) case, the important parameter $\kappa=1-2\xi$, with
\begin{equation}
\xi=\sum_{i=0}^{2}\frac{C_{n-m-1}^{n_i-m}}{C_{n}^{n_{i}}}+\sum_{0\leq i<j\leq 2}\sum_{m_1}\frac{C_{n-m-1}^{n_i+n_j-m}C_{n_i+n_j-m}^{n_j-m_1}C_m^{m_1}}{C_{n}^{n_i}C_{n-n_i}^{n_j}}.
\end{equation}
Where the default value of combination number $C_l^j$ is zero for $j<0$, $l<0$ or $j>l$. The summation on $m_1$ should be limited to $0<m_1<m$. From Figures \ref{Fig.3a} and \ref{Fig.3b}, we can see that the maximal value of $\kappa$ is achieved by comparing the values of $\kappa$ at $(n_2,n_1,n_0)=(n-2,1,1)$ and $(n-1-\lfloor\frac{n-1}{2}\rfloor,\lfloor\frac{n-1}{2}\rfloor,1)$ when we assume $n_0\leq n_1\leq n_2$ without loss of generality.

The results are listed in Table.\ref{T2}.
\begin{table}[h!]
\begin{center}
\caption{Parameters of triseparable condition.}
\label{T2}
\begin{tabular}{l|c|c} 
\hline
\text{m} & \text{n} & \text{n}\\
 & $n_2=n-1-\lfloor \frac{n-1}{2} \rfloor$ & $n_2=n-2$ \\
  & $n_1=\lfloor \frac{n-1}{2} \rfloor,n_0=1$ & $n_1=n_0=1$\\
\hline
1 &  & 4-10\\
3 & & 11\\
4 & 12-15&16-24 \\
5 & 25-33&34-50\\
6 &51-68&69-102\\
7 &103-136&137-207\\
8 &208-269&270-419\\
9 &420-531&532-847\\
10 &848-1029&\\
\hline
\end{tabular}
\end{center}
\end{table}

\begin{figure}
\centering
\subfigure[\label{Fig.4a}]{
\includegraphics[width=1.65in]{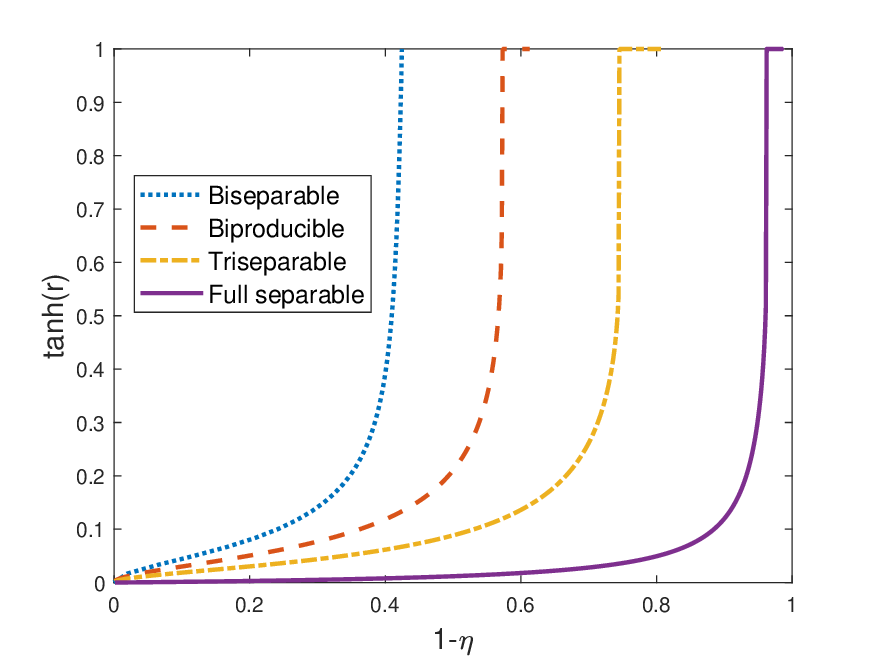}}
\subfigure[\label{Fig.4b}]{
\includegraphics[width=1.65in]{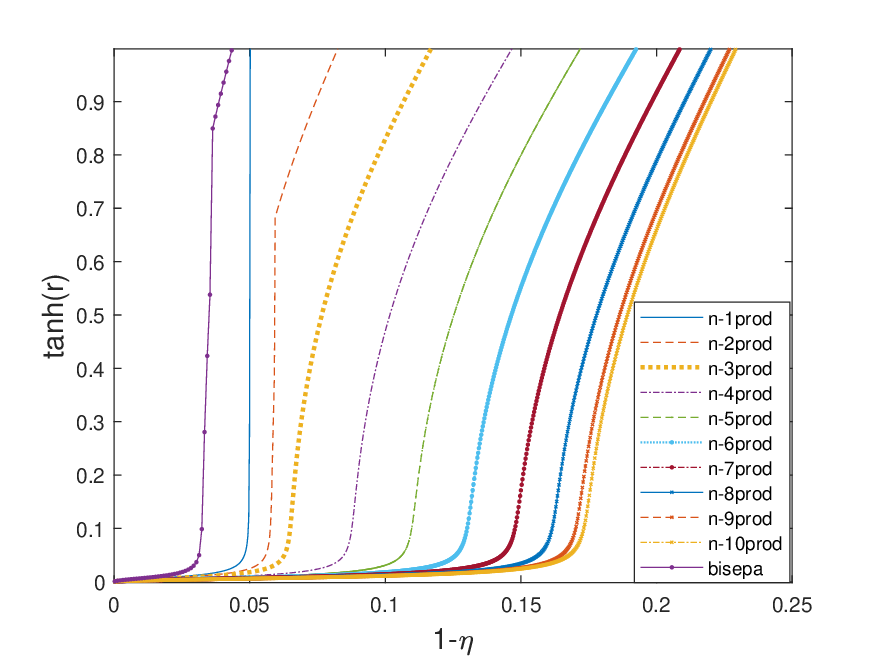}}
\subfigure[\label{Fig.4c}]{
\includegraphics[width=1.65in]{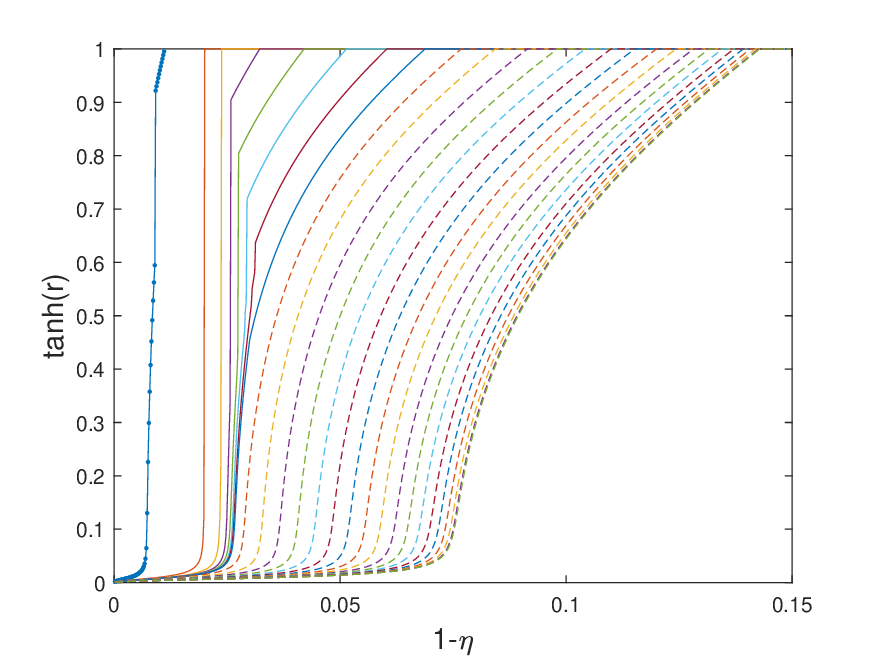}}
\subfigure[\label{Fig.4d}]{
\includegraphics[width=1.65in]{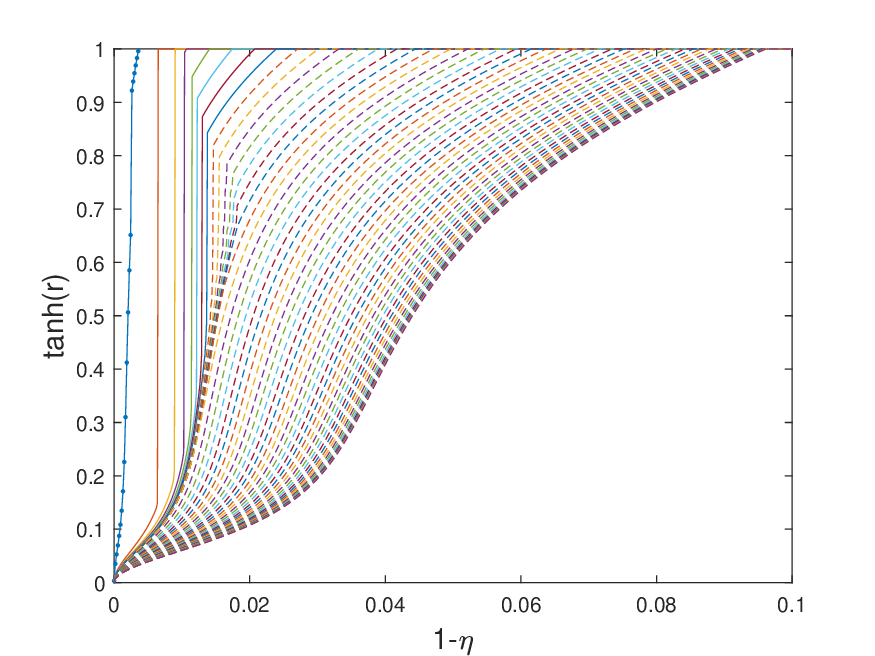}}
\caption{(a).Various necessary conditions of separabilities and producibilities for the four mode CV-GHZ state, $v=0.0001$. (b).$n=20$, lines from from left to right are the biseparable, and verious producible conditions from 19-producibility to 10-producibility, $v=0.0001$. (c).$n=50$, lines from from left to right are the biseparable, and verious producible conditions from 49-producibility to 25-producibility, $v=0.0001$. (d).$n=105$, lines from from left to right are the biseparable, and various producible conditions from 104-producibility to 53-producibility, $v=0.01$.}
\label{Fig4}
\end{figure}

  For the quadripartite separability (quadriseparability) case, we also have $\kappa=1-2\xi$, with
\begin{eqnarray}
&&\xi=\sum_{i=0}^{3}\frac{C_{n-m-1}^{n_i-m}}{C_{n}^{n_{i}}}\nonumber\\
&&+\sum_{0\leq i<j\leq 3}\sum_{m_1}\frac{C_{n-m-1}^{n_i+n_j-m}C_{n_i+n_j-m}^{n_j-m_1}C_m^{m_1}}{C_{n}^{n_i}C_{n-n_i}^{n_j}}\nonumber\\
&&+\sum_{0\leq i<j<l\leq 3}\sum_{m_1,m_2}\frac{C_{n-m-1}^{n_i+n_j+n_l-m}C_{n_i+n_j+n_l-m}^{n_l-m_2}}{C_{n}^{n_i}C_{n-n_i}^{n_j}C_{n-n_i-n_j}^{n_l}}\nonumber\\
&&\times C_{n_i+n_j+m_2-m}^{n_j-m_1}C_{m}^{m_1}C_{m-m_1}^{m_2}.
\end{eqnarray}  
The results are shown in Table.\ref{T3}. Notice that in the descending order of $n_3,n_2,n_1,n_0$, the optimal partition appears at $n_1=n_0=1$. 
\begin{table}[h!]
\begin{center}
\caption{Parameters of quadriseparable condition.}
\label{T3}
\begin{tabular}{l|c|c} 
\hline
\text{m} & \text{n} & \text{n}\\
 & $n_3=n-1-\lfloor \frac{n}{2} \rfloor$ & $n_3=n-3$ \\
 & $n_2=\lfloor\frac{n}{2}\rfloor-1,n_1=n_0=1 $ & $n_2=n_1=n_0=1$\\
\hline
1  &  & 5-10\\
4 & &11-17 \\
5  &18-28&29-38\\
6  &39-63&64-81\\
7 &82-130&131-171\\
8  &172-262&263-357\\
9 &358-523&524-739\\
10 &740-1029  & \\
\hline
\end{tabular}
\end{center}
\end{table}

For $J$-producibility problem of the lossy noisy CV-GHZ state, first of all, we should keep the largest subset of the partition to be $J$-mode. The sizes of other subsets are equal to or less than $J$. If $J\geq\lfloor\frac{n}{2}\rfloor$, a split $\mathcal{I}$ with $J$-size leading subset can be written as $J|n_1|n_2|...|n_l$ with $n_{i}< J$ and $\sum_{i}n_{i}=n-J$. The split $\mathcal{I}$ is the bipartition $J|(n-J)$ itself or its further splits of the $n-J$ part. Clearly the separable state set of $J|n_1|n_2|...|n_l$ is not larger than that of $J|(n-J)$.  Thus we only need to consider the bipartition $J|(n-J)$ instead of all splits $\mathcal{I}$ with $J$-size leading subset. Thus the convex hull of separable state sets reduces to all different bipartitions $J|(n-J)$. Due to the symmetry of lossy noisy CV-GHZ state, all the bipartitions with leading $J$-size subset are equal. Hence for a given $m$ (corresponding to a witness), we can directly calculate the producible condition. Then we compare all the $m$ to find the largest range of non J-producibility. 

The $J$-producible necessary conditions are shown in Figure \ref{Fig4} for 4-mode, 20-mode, 50-mode and 105-mode lossy noisy CV-GHZ states. Where the curves $J$-producibility can be classified as two kinds. The first kind is achieved at $m=1$, the second kind is achieved at $m\neq 1$.  Our calculation shows that when $n\leq 8$, we only have the first kind $J$-producibility. The $n-1$-producibility with $m\neq 1$ apepears when $n\geq 9$. The $n-i$-producibility with $m\neq 1$ appears when $n\geq15,22,28,34,41,47,54,60,66,73,79,86,92,98,105$ for $i=2,3,...,16$, respectively. 

We also list the necessary biseparable conditions for comparison. The biseparable ranges are larger than the $(n-1)$-producible ranges in all the systems except 4-mode system (where the necessary biseparable condition is also the necessary 3-producible condition). Hence a non $(n-1)$-producible state may (with large probability for large $n$) not be genuine entangled. The reason is apperant, a biseparable state can be $(n-1)|1$ split and all the other bipartite splits with leading subset size less than $n-1$, while an $(n-1)$-producible state has to contain $(n-1)|1$ split. 

Notice that the $m=1$ case also appear in the biseparable conditions when the squeezing parameter $r$ is very large, as shown in the upper parts of biseparable conditions in Figures \ref{Fig.4b},\ref{Fig.4c}, \ref{Fig.4d}. For simplicity (and also the conditions are necessary of biseparable), we do not show the $m=1$ improvements of the biseparable conditions in figures of main text when the squeezing parameter $r$ is very large. 

\end{document}